\documentclass[aps, prx, reprint, superscriptaddress, showkeys, amsmath, amssymb,]{revtex4-2}
\usepackage{graphicx}
\usepackage{float}
\graphicspath{ {./Figures/} }

\begin{document}

\title{Back action evasion in optical lever detection}
\date{March 5, 2023}
\author{Shan Hao}
\author{Thomas Purdy}
\email{tpp9@pitt.edu}
\affiliation{Department of Physics and Astronomy, University of Pittsburgh, Pittsburgh, Pennsylvania 15260, USA}

\begin{abstract}
The optical lever is a centuries old and widely-used detection technique employed in applications ranging from consumer products, industrial sensors to precision force microscopes used in scientific research. However, despite the long history, its quantum limits have yet to be explored. In general, any precision optical measurement is accompanied by optical force induced disturbance to the measured object (termed as back action) leading to a standard quantum limit(SQL). Here we give a simple ray optics description of how such back action can be evaded in optical lever detection to beat SQL. We perform a proof-of-principle experiment demonstrating the mechanism of back action evasion in the classical regime, by developing a lens system that cancels extra tilting of the reflected light off a silicon nitride membrane mechanical resonator caused by laser-pointing-noise-induced optical torques. We achieve a readout noise floor two orders of magnitude lower than the SQL, corresponding to an effective optomechanical cooperativity of $100$ without the need for an optical cavity. As the state-of-the-art ultra low dissipation optomechanical systems relevant for quantum sensing are rapidly approaching the level where quantum noise dominates, simple and widely applicable back action evading protocols will be crucial for pushing beyond quantum limits.
\end{abstract}
\maketitle

Optical lever detection, which measures the angular deviation of light reflecting from a tilting surface, is among the oldest precision optical measurement techniques~\cite{Jones1961}. For multiple centuries, the technique has been widely employed in numerous applications ranging from consumer products~\cite{Hsiao2006, Cuomo1989}, industrial sensors such as optical comparators to high precision scientific instruments such as atomic force microscopy(AFM) and its various variants, torsional balances for big-G measurement~\cite{Rothleitner2017}, and active mirror alignment for gravitational wave detectors~\cite{Hirose2010,Keiko2018}. Although the optical lever is old and simple, contrary to general intuition, it is as sensitive as optical interferometry~\cite{RN65,Barnett2003}, which has been more associated with precision measurement. However, the quantum limits of optical level detection have yet to be experimentally explored, as systems to date have been limited by thermal or other classical noise sources. Advances in nanomechanical resonators in the past few years ~\cite{Ghadimi2018, Erick2020, Mohammad2021, Tsaturyan2017, MacCabe2020, Pratt2021}have pushed the device quality factor to a new level reducing thermal noise, making quantum noise caused by optical forces (i.e., measurement back action) a dominant limit to be overcome. Here, we show that the effects of back action are evaded by modifying the optical lever beam path with a carefully designed lens system.

In optical lever detection, quantum measurement back action presents as noisy torques from random arriving photons (shot noise) recoiling off a device. The standard quantum limit(SQL) is the minimum noise achieved by balancing shot noise and its back action induced disturbance. Methods that sidestep this limit~\cite{Braginsky1992, Kimble2001, Clerk2010} are referred to as back action evasion. So far, the SQL and protocols for beating it have been extensively studied in interferometric optomechanical systems~\cite{Suh2014, Mason2019, Yu2020}. Surpassing the SQL will extend the scope and sensitivity of gravitational wave detectors~\cite{LIGO2016} and enhance searches for dark matter candidates~\cite{Carney2020,Carney2021} and tests of the validity of quantum mechanics at a macroscopic scale~\cite{Carlesso2022}. Most of the work done so far on back action evasion utilizes optical cavities to enhance optomechanical interaction strength. On the other hand, a cavity limits the bandwidth, complicates the setup, and makes the system prone to classical laser noise. One notable exception is recent work in levitated nanoparticle systems that approach the regime where back action can be evaded without using a cavity~\cite{militaru2022, Magrini2022}.  

The typical angular resolution for optical lever detection is on the order of $\mathrm{10\,prad/\sqrt{Hz}}$ for applications such as AFMs~\cite{Alexander1989, Gerhard1988}. The technique has been pushed further with multiple bounces~\cite{Hogan2011}, non-classical laser sources such as single or multimode squeezed light~\cite{Treps2003, Pooser2015}, and with cavity enhancement\cite{Komori2020, Shimoda2022}. However, these experiments are still far from the SQL. A method to surpass SQL was theoretically proposed only a few years ago by Enomoto, et al.~\cite{Enomoto2016}. Noise correlations introduced by the optomechanical interaction produce squeezing (sometimes referred to as ponderomotive squeezing) and can be used to evade back action in an optical lever detector. We present a simple ray optics picture of this process and experimentally demonstrate this protocol for back action evasion at room temperature with classically driven noise. We observe an optomechanical cooperativity, which characterizes measurement strength relative to the SQL, up to $100$. We also find that parameters needed for evading quantum back-action are within experimental reach. This quantum-enhanced optical lever detection protocol will be beneficial for ultra-high precision measurements such as quantum AFM or scanning force microscopy utilizing high Q mechanical resonators~\cite{Halg2021} and novel types of gravitational-wave detectors~\cite{Ando2010}.

\section{Quantum noise in optical lever}
\begin{figure}[ht]
\includegraphics[scale = 0.245]{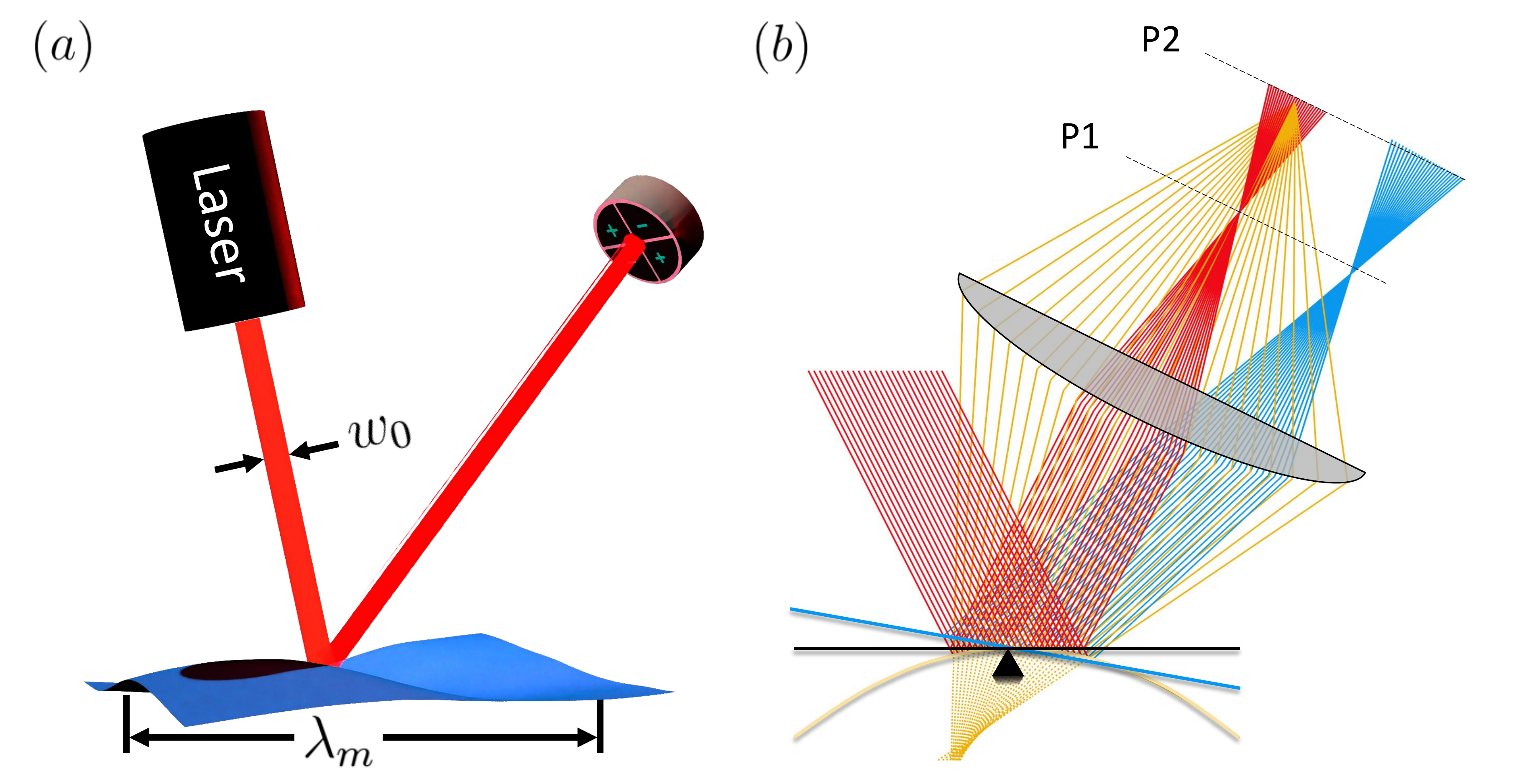}
\caption{\label{fig:HGexpansion} \textbf{Visualization of back action evasion in optical lever} (a) Optical lever detection of a membrane mechanical resonator. The laser is reflected by a small area of the membrane. The vibrational motion of the membrane tilts that small area, and deviates the laser spot from its balanced position at the split photodetector. (b) Ray optics visualization of back action evasion in the low frequency limit. If the surface is fixed, the incoming red beams are reflected and focused by a lens to one point. If the mirror tilts due to external force, the rays will be focused to a different point (blue lines) on plane P1(Fourier plane). The yellow rays take account of the result of optical forces on the mirror. Rays impinging further from the pivot point give larger back action torques, and so the yellow rays appear to be reflected off a parabolic mirror and focused to plane P2, which we term back action evasion plane. The mirror curvature, and thus the location of P2, is frequency dependent following the mechanical response function. The maximum SNR sits between P1 and P2.}
\end{figure}

An optical lever, in itself, is very simple as shown in Fig.~\ref{fig:HGexpansion}(a). In our system, a laser reflects off the node of a suspended membrane mechanical resonator. Under the condition that the laser beam waist ($w_0$) is much smaller than the mechanical wavelength ($\lambda_m$), the out-of-plane motion of the membrane at the node can be approximated as a tilting mirror. The deflection of the laser is then measured by a split photodetector, which takes the difference of the optical power on left and right sides (see the Supplemental Material for details). The precision of optical lever detection is limited by quantum noise, as photons strike the split photodetector randomly in time and with a spacial distribution following the beam intensity profile (represented by finite beam width in Fig.~\ref{fig:HGexpansion}(a) and spread rays in Fig.~\ref{fig:HGexpansion}(b)). This shot noise limited readout floor improves with laser power. Additionally, the random recoil of photons off the membrane imposes an imbalanced force noise, which tilts the mirror and redistributes the reflected photons. Such back action noise increases with laser power. The standard quantum limit resides where the sum of the two noise sources is at a minimum.

A simple case of back action evasion in an optical lever can be visualized in a ray optics picture as shown in Fig.~\ref{fig:HGexpansion}(b). Uniformly distributed incoming rays reflect off a flat mirror and are focused through an output lens to its Fourier plane, P1 (red rays). Tilting of the mirror (blue line) by an external force displaces the focal point off the optical axis (blue rays). An incoming photon's beam path is described by a random incoming red ray. Considering such a single ray off the axis, it gives a torque proportional to arm length, such tilting of the mirror gets larger for the ray further off the pivot point and thus deflect it further from the original reflection direction. Thus, the beam path for a given ray is equivalent to reflecting off a convex parabolic mirror (yellow line). In other words, the flat mirror effectively picks up a curvature due to such back action. The reflected photons (yellow rays) can be made to converge to the optical axis at a certain distance beyond P1. A split photodetector placed there thus sees no effects of back action, but retains sensitivity to the tilting of the mirror (blue line) from external forces.

\section{Gaussian optics picture}

To better quantify the effects of quantum noise, we decompose the optical field in the basis of Hermite-Gaussian (HG) modes, generalizing the treatment of~\cite{Enomoto2016} (see the Supplement material \cite{supp} for details). $U_{mn}$ is the HG mode of order $m$ in $x$ direction and $n$ in $y$ direction. A laser in a coherent state of the $U_{00}$ mode reflecting from an undulating surface can be thought of as scattering light into higher order HG modes.  To first order, a small tilt of this Gaussian laser beam can be characterized as the addition of a $U_{10}$ component with a small imaginary amplitude (in addition, small transverse displacements of the laser correspond to a real amplitude $U_{10}$ component).  This limit is relevant for our case where $w_0 \ll \lambda_m$ and the amplitude of motion of the membrane is small, and forms the basis of optical lever measurement. 

In this case, the input field can be approximated as~\cite{Morrison1994, Wunsche2004} 
\begin{equation}
\alpha U_{00} + (d \hat{X}_1+ id \hat{X}_2)U_{10},\label{HGQ}
\end{equation}
\noindent where $\alpha \gg 1$ is the coherent amplitude. $d \hat{X}_1 $ and $d \hat{X}_2$ are vacuum amplitude and phase fluctuations of $U_{10}$ that give rise to quantum limited beam displacement and pointing noise of the laser, respectively. The latter is responsible for readout noise at the detector, whereas the former seeds the mechanical back action on the membrane. The reflection due to small back action induced tilting imprints amplitude noise of $U_{10}$ onto its phase quadrature as (see the Appendix):

\begin{eqnarray}
&d \hat{X}_{1}(\omega) \rightarrow d \hat{X}_{1}(\omega)\label{corr1}\\
&d \hat{X}_{2}(\omega) \rightarrow d \hat{X}_{2}(\omega) + 2\hbar D\chi_m(\omega)\cdot d \hat{X}_{1}(\omega).\label{corr2}
\end{eqnarray}

\noindent Here $D = N|kk_mw_0|^2$ and $\chi_m(\omega)$ is the mechanical response function. $N = |\alpha|^2$ is the photon flux, $k$ is the laser wave number and $k_m = \frac{2\pi}{\lambda_m}$. The generated correlation between amplitude and phase quadrature of the reflected beam is very similar to ponderomotive squeezing in cavity optomechanics~\cite{Brooks2012, Purdy2013, Safavi2013} but different in that another spacial mode is put in a squeezed vacuum state as opposed to being bright squeezed. 

$U_{00}$ does not contain any information about the tilting of the membrane, but it does act as a local oscillator when multiplied with the split photodetector step weighting function ($\pm 1$ for two halves), to extract information from $U_{10}$. As the reflected beam propagates, the Gouy angle $\theta(z)$ generates a relative phase shift of $\theta$ between $U_{00}$ and $U_{10}$. Thus, the Gouy angle is equivalent to the measured quadrature of $U_{10}$ (see the Appendix B for further details). Picking a particular quadrature just requires putting the split photodetector at a specific location along the beam path, or adding lenses to adjust the Gouy phase.

Putting everything together, we can write down the expression for power spectrum of the measured membrane tilt angle $\beta$ as

\begin{equation}
S_{\beta} = S_{\beta}^{Ba} + S_{\beta}^{Imp} + S^{Ba, Imp}_{\beta} + S_{\beta}^{Th} + S_{\beta}^{Detc},\label{SXX}
\end{equation}

\noindent where $S_{\beta}^{Ba} = N\hbar^2 k^2 w_0^2 k_m^4 |\chi_m|^2$ is the back action of the laser on the membrane, $S_{\beta}^{Imp} = \frac{1}{4 N w_0^2 k^2 \sin^2\theta}$ is the imprecision readout noise from the laser shot noise, $S_{\beta}^{Th}$ is the thermal and zero point motion of the membrane, $S_{\beta}^{Detc}$ counts for the loss of quantum efficiency in detection, and $S^{Ba, Imp}_{\beta}=\hbar k_m^2\, \cot(\theta)\, \mathrm{Re}\{\chi_m\}$ is the interference term from the optomechanically induced correlation between the two quadratures described by Eq.~(\ref{corr1}) and Eq.~(\ref{corr2}) (see the Appendix B for details). It can be shown that $ S^{SQL}_{\beta} = \hbar k_m^2 |\chi_m|$, which is the best one can achieve neglecting the last 3 terms in Eq.~(\ref{SXX}), and with the correlation $S_{\beta}\geq \hbar k_m^2 \mathrm{Im}\{\chi_m\}$ (see the Appendix B for details). The SQL can be beaten at large optical power with low thermal and detection noise, if a good combination of $\theta$ and $\omega$ is found to maximize the destructive interference term $S^{Ba, Imp}_{\beta}$. This explains why back action evasion occurs at a certain distance from the device.

\section{Proof-of-principle experiment}

\begin{figure}
\includegraphics[scale = 0.28]{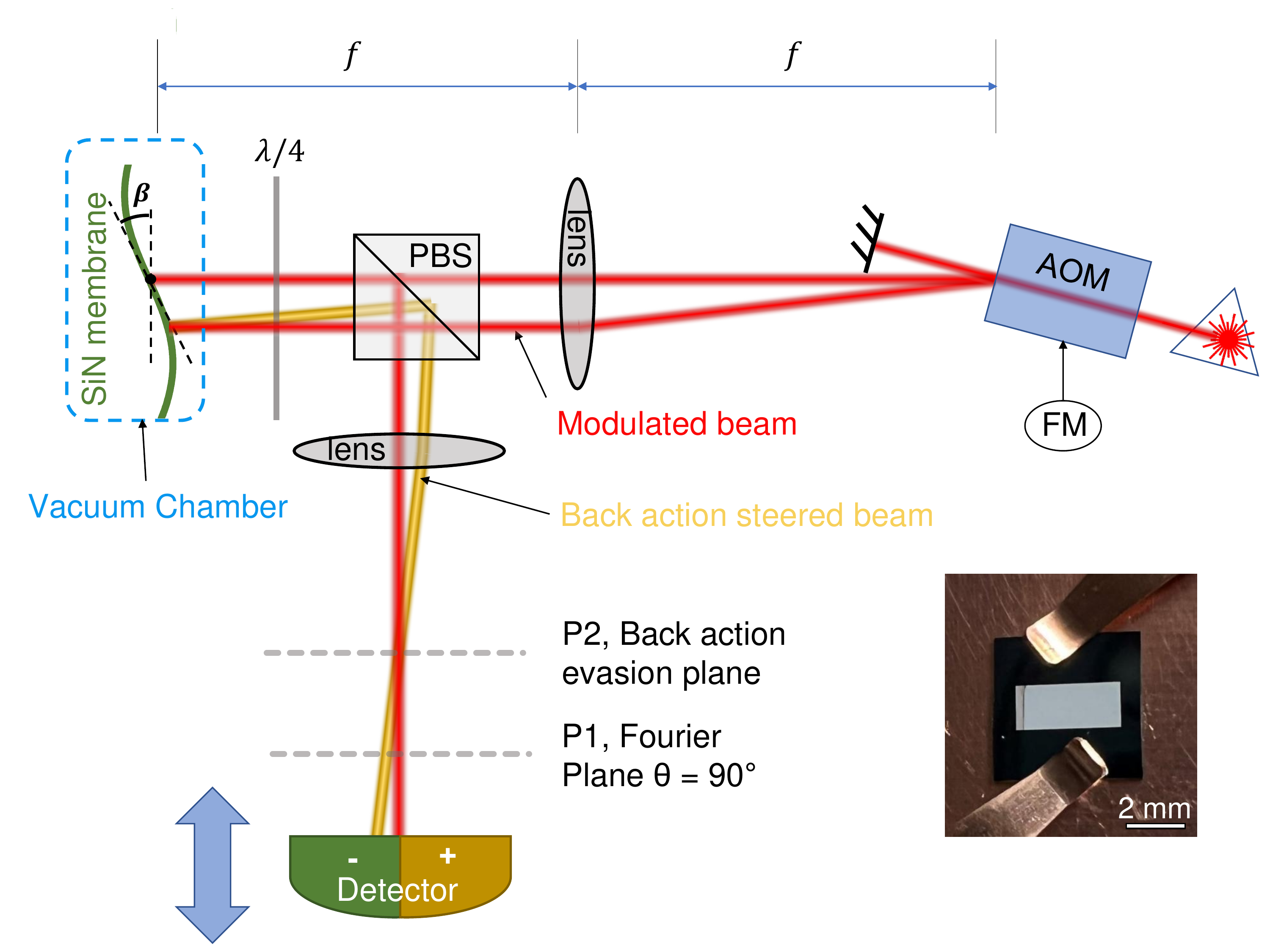}
\caption{\label{fig:EXP} \textbf{Experiment scheme.} An FM modulated AOM is placed at one focal plane of the input lens and the SiN membrane is at the other, so that the angular modulation of the first order diffracted beam turns into position modulation at the plane of the device. The modulated beam disturbs the motion of the membrane and deflects itself into a different direction (yellow). The illustrated case corresponds to a modulation frequency higher than the mechanical resonance frequency, so that the mechanical response lags $180^{\circ}$ behind the modulated force, leading to the reflected beam being steered toward the undisplaced beam and putting the back action evasion plane P2 before P1. Inset is a picture of the SiN membrane device clapped on the sample holder.}
\end{figure}

At room temperature, thermal motion dominants over quantum back action for devices we currently have. To demonstrate the principle of back action evasion, we intentionally add classical laser displacement fluctuations by modulating the laser spot position. As shown in Fig.~\ref{fig:EXP}, a commercially available (Norcada Inc.) silicon nitride (SiN) low stress (about 200~MPa) membrane (3.5~mm long, 1.5~mm wide and 100~nm thick) is clamped inside a vacuum chamber (1.3e-7~Torr) and positioned at the focal plane of the objective lens ($f = $~125 mm). An Acousto-Optic Modulator(AOM) is located at the other focal plane of the objective lens and is frequency modulated(FM) so that angular deviation of the 1st order diffracted beam is turned into laser (1064~nm) spot position modulation at the membrane. The reflected laser is then steered through an output lens ($f = $~125 mm) to a split photodetector that can move along the beam path to measure at different quadrature angles. The output lens maps the far field onto its Fourier plane, generating additional accumulated Gouy phase ~\cite{Erden1997} (see the Appendix B for accumulated Gouy phase) allowing us to access all quadratures with finite movement of the split photodetector. Near the Fourier plane, the injected amplitude noise is mostly rotated to the phase quadrature, which the split photodetector is insensitive to. In the meantime, the quadrature that carries information about the vibrations of the membrane is rotated mostly to amplitude quadrature.

We modulate the laser spot position by chirping (which play the role of classical noise) the AOM FM frequency across the resonance of the mechanical mode of interest ($f_m \approx 371 \mathrm{kHz}$). The resultant driven motion is much larger than the thermal motion. Then we measure the spectral response of the tilting angle at different quadrature angles($S_{\beta}(\theta)$) by translating the split photodetector along the beam path near the Fourier plane and averaging at each quadrature angle for longer than the decay time of the mode (see the Supplemental Material for experiment details). 

A density plot of the measured spectra against quadrature angles is shown in Fig.~\ref{fig:BAdata}(b). By modifying Eq.~(\ref{SXX}) to incorporate such classical noise (see the Appendix C for detail), we calculate the corresponding theoretical density plot based on independently measured experimental parameters as shown by Fig.~\ref{fig:BAdata}(a). To visualize the spectra $S_{\beta}(\theta)$, we normalize to the measured background noise floor $S_{\beta}^{Cla}(\theta)$, i.e., the projection of the classical injected noise to the measured quadrature. Such normalization mimics that of the spectrum to laser shot noise in the case of quantum back action evasion, by considering our classical injected noise as artificially amplified quantum amplitude fluctuations.  In the case of pure quantum noise, the shot-noise-limited phase quadrature fluctuations  would contribute a significant imprecision background to the overall signal, creating a quadrature-independent background level.
\begin{figure*}
\includegraphics[scale = 0.65]{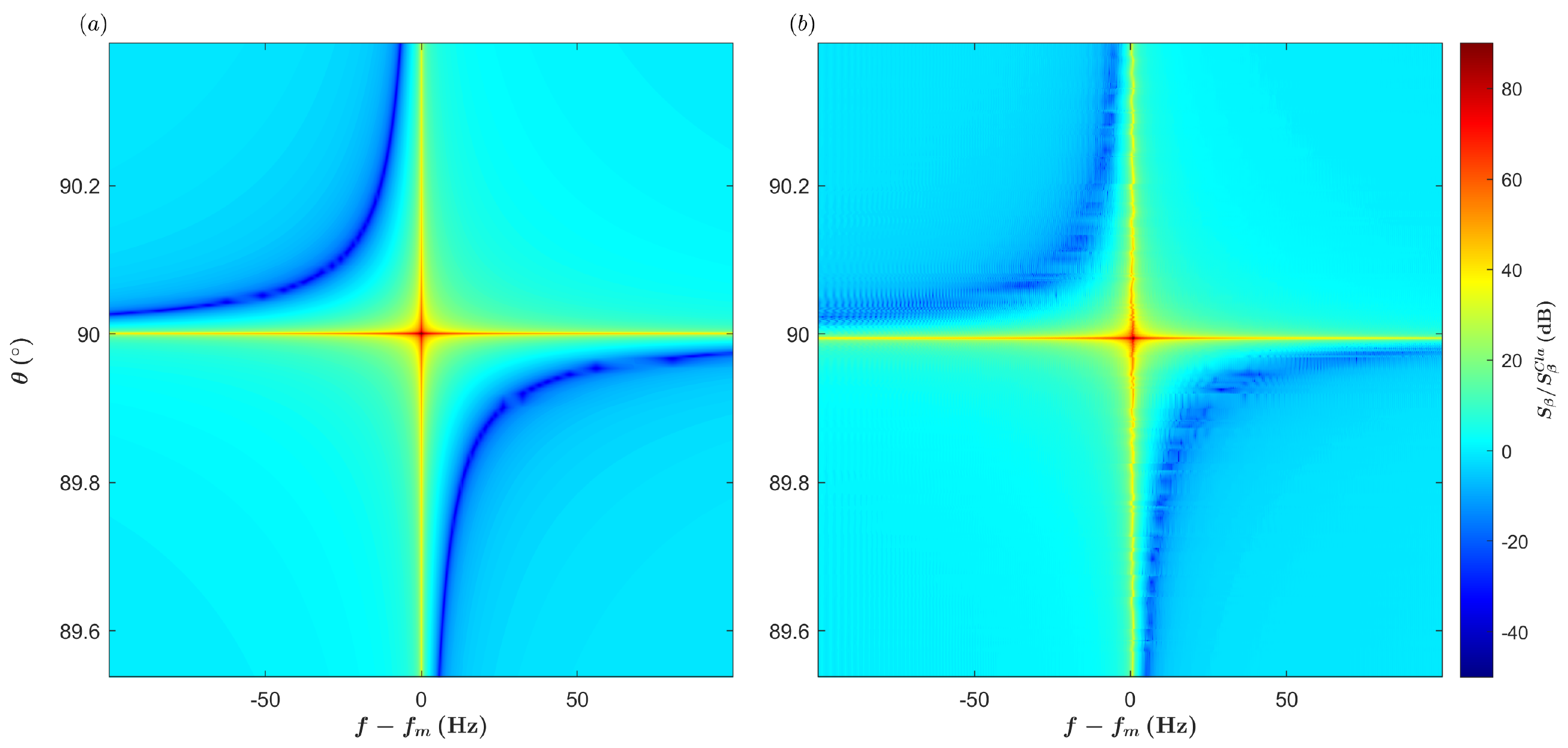}
\includegraphics[scale = 0.65]{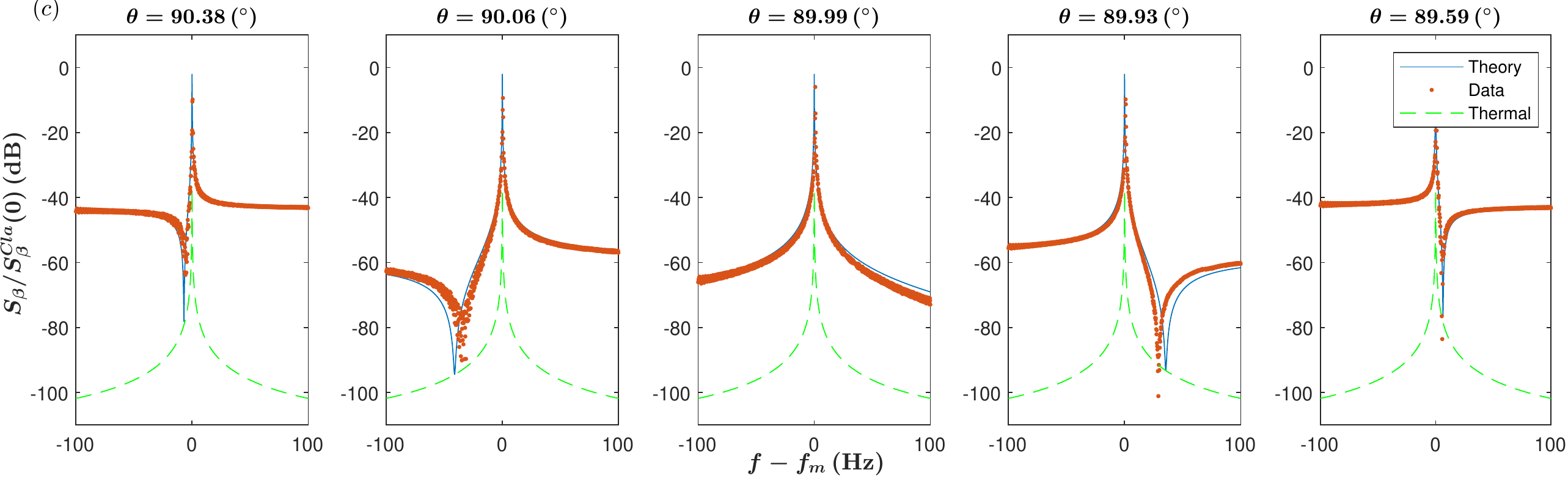}
\caption{\label{fig:BAdata} \textbf{Demonstration of classical back action evasion.} (a) Zero free parameter theoretical expectation of $S_{\beta}$ normalized to injected noise at corresponding quadrature angle $S_{\beta}^{Cla}(\theta)$ versus quadrature angles and frequencies. (b) Corresponding experiment data. Mode frequency $f_m = 371$~kHz, $Q = 3.2\times 10^{6}$, $m_{eff} = 4.06\times 10^{-10}$~kg, $T = 297.5$~K and the laser power $P = 21$~mW. Observation of region with less than 0~dB demonstrates the interference mechanism of back action evasion in optical lever detection. (c) Expectation and experimental data of $S_{\beta}$ at several representative quadrature angles $\theta$ normalized to input noise $S_{\beta}^{Cla}(\theta = 0)$. The curves can be interpreted as the gain function at different quadrature angles. The green dashed curves mark the thermal motion level, which stops dips from going deeper and demonstrates our ability to resolve small motion on top of the large injected noise. Due to the sub-Hz resonance frequency random drift which is beyond our active control capability, the sharp peaks and dips are blurred by random resonance frequency drift. }
\end{figure*}
\noindent  The region below 0~dB is where the perceived motion spectrum is smaller than the artificially injected noise floor, a classical equivalent of ponderomotive squeezing.

Figure~\ref{fig:BAdata}(c) are several spectra from the same data set at some representative quadrature angles. These spectra are normalized to the input noise $S_{\beta}^{Cla}(\theta = 0^{\circ})$. The curves can thus be interpreted as the gain function of the system with the peak of the motion approaching 0~dB, corresponding to a measured cooperativity close to 1. Zero free parameter theoretical expectations are shown as solid blue lines to have excellent qualitative agreement with experiment data. At $90^{\circ}$, the curve is just the Lorentzian of the classical noise driven motion of the mechanical mode with no back action evasion, equivalent to what will be measured in the far field. At other quadrature angles, the destructive interference (the dips in the curves) of injected noise and its driven motion emerges in a frequency and quadrature dependent fashion, demonstrating the mechanism of back action evasion in optical lever detection. Green dashed lines are the expected thermal motion spectrum, which stops the back action evasion (the dips) from reaching deeper, and demonstrates our ability to resolve small motion on top of the large injected noise.

\section{Cooperativity in optical lever}

\begin{figure}
\centering
\includegraphics[scale = 0.43]{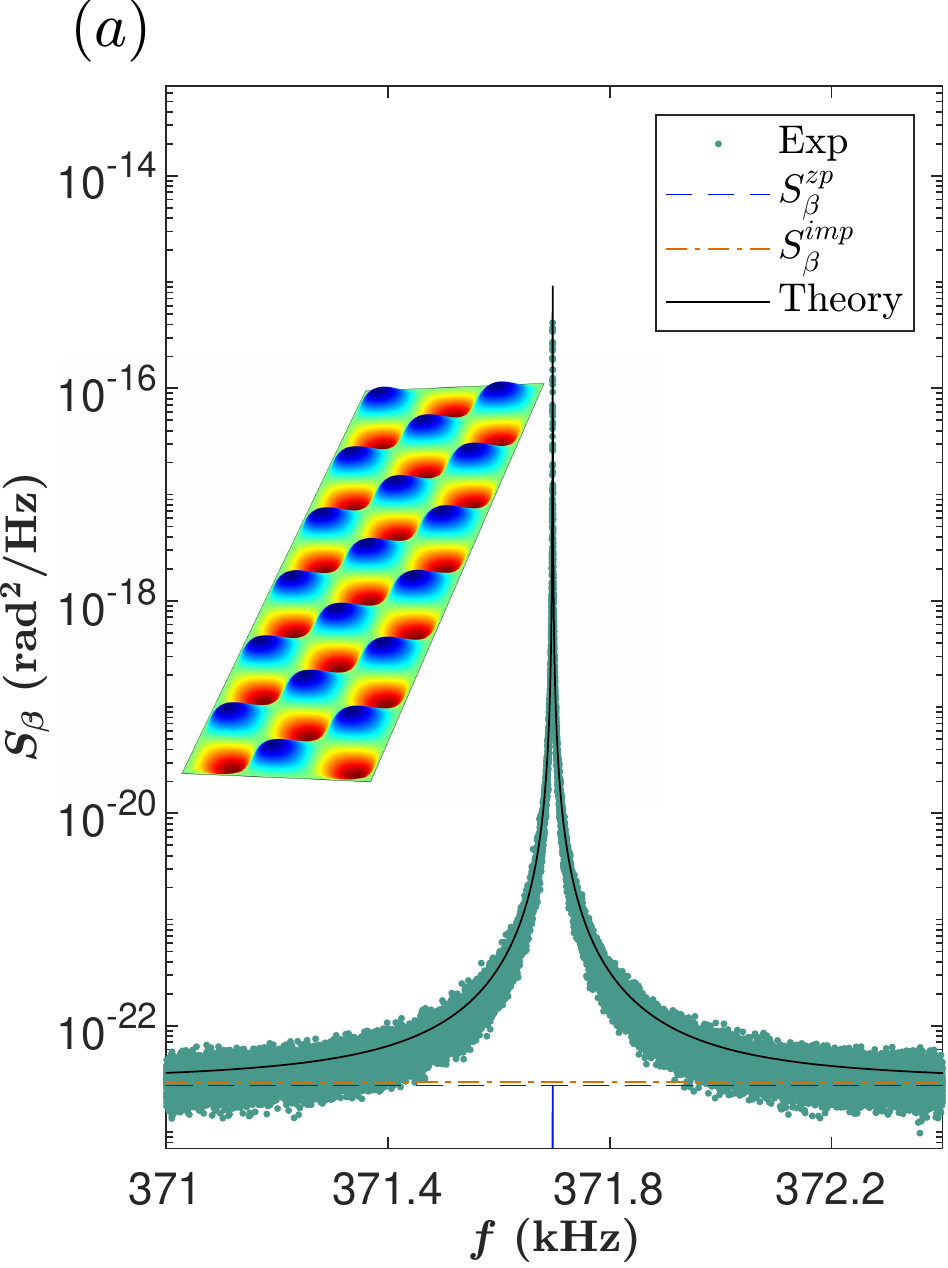}
\includegraphics[scale = 0.43]{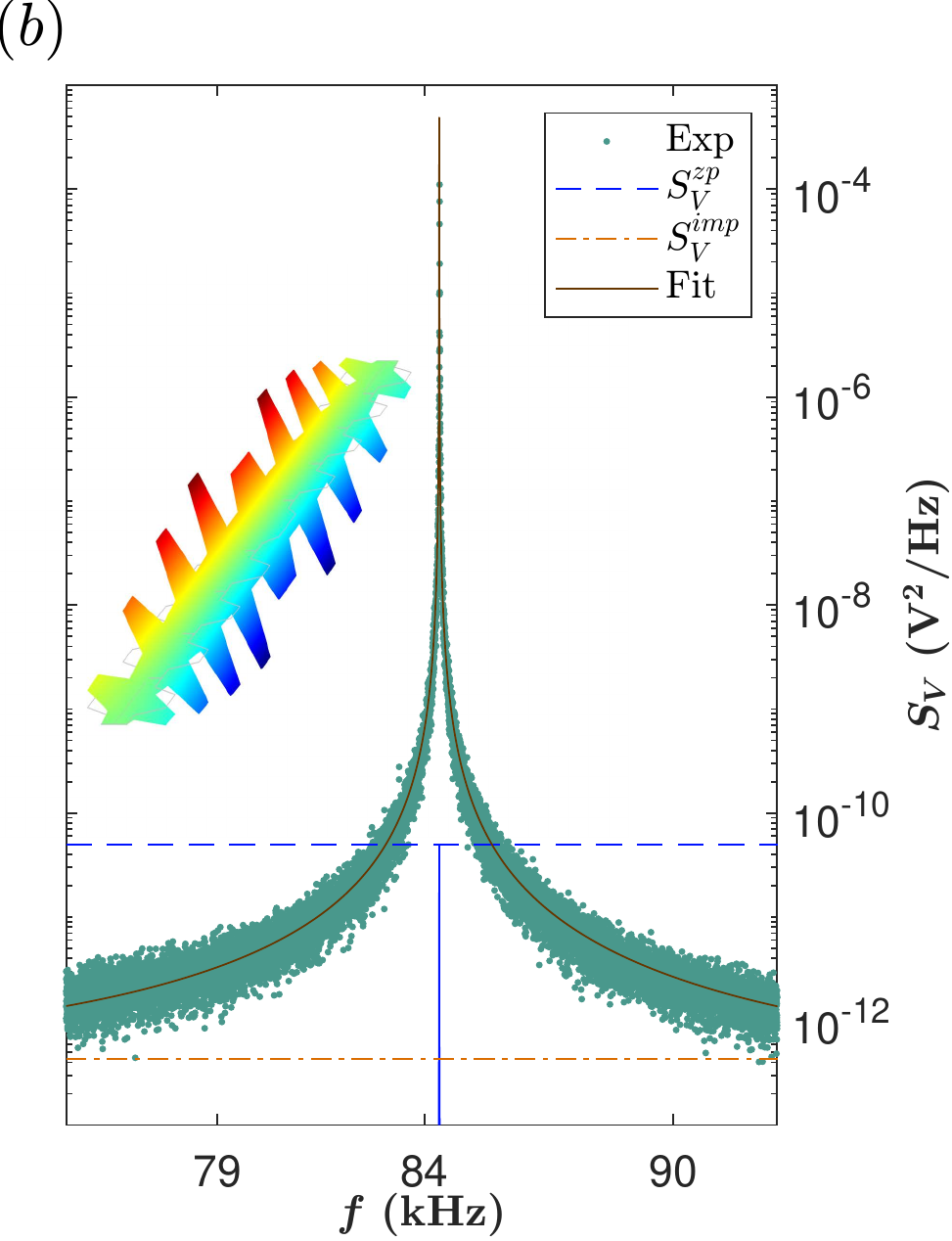}
\caption{\label{fig:coop} \textbf{High cooperativity optical lever measurements.} (a) Room temperature thermal motion PSD of the membrane device inspected in back action experiment. The laser power is instead $P = 27.2$~mW. The green dots are experimental data averaged for over 30 minutes. Black is the zero free parameter calculation. The blue dashed line marks the peak of zero point motion of the device, which is $\mathrm{5.23\,prad/\sqrt{Hz}}$. The orange dash-dotted line marks the measured imprecision noise level, which is $\mathrm{5.45\,prad/\sqrt{Hz}}$, slightly above the zero point motion. (b) Thermal motion PSD (as signal $V$ measured by a split photodetector) of a SiN phononic crystal string of $f_m \approx 84$~kHz, $Q \approx 1.0\times 10^{5}$ and $m_{eff} \approx 7.3\times 10^{-13}$~kg at $T = 20$~K. A Lorentzian fit yield the cooperativity to be approximately $C \approx 100$. Insets show mode shapes from simulation.} 
\end{figure}

Quantum back action evasion requires large optomechanical coopperativity, $C$. For an ideal measurement, a cooperativity of 1 indicates the ability to resolve signals at the scale of the zero point motion and the regime where the back action induced motion becomes comparable to the zero point motion. Derived from the signal-to-noise ratio (SNR) of zero point motion to shot noise limited readout floor, the explicit form of the effective cooperativity for a mechanical membrane or string of sinusoidal mode shape and under small beam waist approximation can be written as 

\begin{equation}
    C = \frac{2PQ}{m_{eff}}\frac{kw_0^2k_m^2}{c\omega_m^2},\label{C}
\end{equation}

\noindent where $P$ is the laser power reflected from the device, $Q$ is the mode quality factor, $m_{eff}$ is the effective mass of the mode, and $\omega_m$ is the mechanical mode angular frequency. 

Figure~\ref{fig:coop}(a) shows the measured thermal motion spectrum of the mode probed in the back action evasion experiment (see Supplement materials \cite{supp} for details regarding experiment calibration). The zero-point motion spectrum, $S_{\beta}^{zp}$, is at the same level as $S_{\beta}^{imp}$. From Eq.~(\ref{C}) and with the measured experiment parameters, we estimate the cooperativity to be $C \approx 2$. Given the estimated net quantum efficiency $\eta \approx 21\%$, our measured SNR of the thermal motion spectrum (peak to noise floor) is consistent with the theoretical expectation ($\mathrm{SNR} = 4\eta n_{th} C$, $n_{th}$ is the thermal phonon number). 

There are several factors contributing to the finite quantum efficiency. Firstly, an ideal split photodetection measurement, despite only detecting whether the photons fall on the left or right half, has a surprising large 64\% quantum efficiency~\cite{Treps2002, Knee2015, Walborn2020}. Secondly, the approximation that the beam waist ($w_0\approx\mathrm{100~\mu m}$) is much smaller than the mechanical wavelength ($\lambda_m\approx\mathrm{580~\mu m}$) is only partially satisfied. This complicates the reflection process at the device by scattering photons into higher HG modes ($m\geq 3$) and still producing back action. At the split photodetector, higher order HG modes contribute to the signal with even lower quantum efficiency and tend to interferometrically cancel due to relative Gouy phase shift (see the Supplemental Material regarding Gouy phase shift in lens system). Lastly, absorption and losses along the beam path and the photodetector conversion efficiency amount to 87\% quantum efficiency.

The theory curves in Fig.~\ref{fig:coop} have already taken into account the various factors discussed above (also for theory curves in Fig.~\ref{fig:BAdata}), and agree very well with experiment data. Still, even after we actively stabilize the thermal drift of resonance frequencies (see the Supplemental Material for details regarding frequency stabilization), we observe random drift of the resonance frequency at sub-Hz level, precluding us from fully resolving the tip of the peak. To illustrate the achievability of large cooperativity in an optical lever system and to push the experiment toward quantum back action evasion, we measure the thermal motion power spectrum density (PSD) of a SiN phononic crystal string~\cite{Ghadimi2018} (about 2~mm long, $\mathrm{10~\mu m}$ wide and 120~nm thick) fundamental torsional mode with laser of power $P = \mathrm{12.2~mW}$ focused to a beam waist $w_0 = \mathrm{4 ~\mu m}$ at $T = 20$~K (see Supplement material \cite{supp} regarding laser heating and the estimate of device temperature.), as shown in Fig.~\ref{fig:coop}(b). Using $Q$ measured from ring down, we fit the Lorentzian PSD and estimate the cooperativity to be around $C\approx 100$. Compared to the membrane, the cooperativity of the string is much higher due to its smaller effective mass $m_{eff}$. We also note a recent report~\cite{Pratt2021} of higher SNR for torsional mode with wider strings of much higher $Q$.

\section{Conclusion and Outlook}
In this work, we revisited the two-hundred-year-old optical lever technique for reaching and beating the SQL. We have detailed the physics of back action evasion in optical lever detection, which can be beneficial for precision measurement applications. Our proof-of-principle demonstration of classical noise evasion can already be applied for typical optical lever detection that suffers from classical ambient noise. We achieve an optomechanical cooperativity $C > 1$ for the membrane device inspected, which agrees well with theory. Also, benefiting from the lower effective mass of torsional mode of the string, we found a cooperativity $C \sim 10^2$. 

To observe quantum back action evasion, we will need the cooperativity to be larger than the thermal occupation. This is possible with parameters available from state-of-the-art optomechanical devices. With phononical crystal shielding, soft clamping~\cite{Fedorov2020, Mohammad2022}, and dissipation dilution~\cite{Ghadimi2018}, effective masses $m_{eff}$ can be made much smaller and quality factors can exceed $Q > 10^9$. Strained crystalline material such as Si~\cite{Beccari2022} or SiC can still have lower dissipation at cryogenic temperature. 2D resonators have better laser power handling and have been demonstrated to have comparable quality factors as that of strings~\cite{Tsaturyan2017, Shin2022}. Photonic crystal mirrors can also be utilized to push reflectivity close to unity~\cite{Lawall2012, Norte2016} to increase reflected laser power and potentially incorporating multiple bounces~\cite{Hogan2011, Tomofumi2022}. With these improvements, quantum back action will be the dominant noise at cryogenic temperature, but can now be readily dealt with.

\bibliography{main}
\section*{Acknowledgement}
This work is supported by NSF foundation under Award No.2047823. We acknowledge support from the Charles E. Kaufman foundation. We thank Robinjeet Singh for fabricating our string mechanical resonator at the NIST Center for Nanoscale Science and Technology.

\appendix
\section{\uppercase{Optical field of the laser}}
The input field of a typical laser of power $P$ in the frame rotating at laser frequency $\omega_L$  can be written in the basis of Hermite-Gaussian(HG) modes as $$\alpha U_{00} +\sum_{mn}\hat{\delta}_{mn}U_{mn}. $$
Here $U_{mn}(x,y,z)$ are $m$th order in $x$ and $n$th order in $y$ HG mode functions for modes propagating along z and focused to waist $w_0$ at $z = 0$ where membrane is located (see the Supplemental Material\cite{supp} for detailed definition of HG beam). $|\alpha|^2 = N = \frac{P}{\hbar\omega_L}$ and $\alpha$ is assumed to be real without loss of generality, and $\hat{\delta}_{mn}$ represents the vacuum fluctuation in $U_{mn}$.  $\hat{\delta}_{mn} = d\hat{X}_1^{mn}+id\hat{X}_2^{mn}$ are quantum Langevin noise operators where $d\hat{X}_1^{mn}$ is the amplitude vacuum fluctuation, $d\hat{X}_2^{mn}$ is the phase vacuum fluctuation and $\langle d\hat{X}_i^{mn}(\omega) (d\hat{X}_j^{pq}(\omega^{\prime}))^{\dagger} \rangle = \frac{1}{4}\delta_{ij}\delta_{mp}\delta_{nq}\delta(\omega - \omega^{\prime})$

\section{\uppercase{Fourier optics calculation}}

The reflection of the laser field off the device can be calculated with Fourier optics. The reflected field $\hat{U}_{out}$ can be associated with the input field $\hat{U}_{in}$ as (up to a change of direction of propagation)$$\hat{U}_{out}(z=0) = \hat{U}_{in}e^{-i2k\hat{\Phi}(\vec{r},t)},$$ where $k = \frac{\omega_L}{c}$,  $\hat{\Phi}(\vec{r},t) = \hat{X}(t)\phi(\vec{r})$, $\phi(\vec{r})= \sin(\frac{l\pi}{a}(x-\frac{a}{2}))\sin(\frac{p\pi}{b}(y-\frac{b}{2}))$ is the out-of-plane motion mode shape (origin put at the center of the membrane, $a$ and $b$ are membrane side lengths, and $l$ and $p$ are mechanical mode indices) and can be approximated to $\hat{\Phi}(\vec{r},t) \approx \hat{X}(t)k_mx$ as the laser beam waist is much smaller than the mechanical wave length ($w_0\ll \frac{2\pi}{k_m}$, $k_{m}$ is mechanical wave numbers) and centered to the anti-node in y direction and the node in the x direction (center of the membrane). $\hat{X}(t)$ satisfy the equation of motion, $$m_{eff}\ddot{\hat{X}}+m_{eff}\Gamma\dot{\hat{X}}+m_{eff}\omega_m^2\hat{X} = \hat{F},$$
where $\hat{F}$ is the force acting on the membrane and $\Gamma$ is the damping rate. In the frequency space
\begin{equation}
\hat{X}(\omega) = \chi_m(\omega)\hat{F}(\omega),\label{EOM}
\end{equation} where $\chi_m(\omega) = \frac{1}{m_{eff}}\frac{1}{(\omega_m^2-\omega^2)+i\omega\Gamma}$.

With the approximation made above, we have
\begin{align*}
    \hat{U}_{out}(z=0) &\approx \hat{U}_{in}(1-i2k\hat{X}(t)k_mx)\\
            &\approx (\alpha U_{00}+\sum_{mn}\hat{\delta}_{mn}U_{mn})(1-i2k\hat{\beta}(t)x)\\
            &\approx \alpha U_{00}+(d\hat{X}_1^{10}+i(d\hat{X}_2^{10} - \alpha w_0 k \hat{\beta}(t)))U_{10}\\
            &+ \sum_{mn\neq 10}\hat{\delta}_{mn}U_{mn},
\end{align*}
where we have used the fact that $\frac{2}{w_0}xU_{00} = U_{10} $ and $\hat{\beta}(t) \equiv \hat{X}(t)k_m$ is the tilting angle at the node.

Propagating the reflected field out, we have the light field at distance as:
\begin{align*}
    \hat{U}_{out}(z) &= \alpha U_{00}(x,y,z)\\
    &+(d\hat{X}_1^{10}+i(d\hat{X}_2^{10} - \alpha w_0 k \hat{\beta}(t)))U_{10}(x,y,z)\\
    &+ \sum_{mn\neq 10}\hat{\delta}_{mn}U_{mn}(x,y,z),
\end{align*}

The signal $V(t)$ for the split photodetector at $z$ is proportional to 
\begin{align*}
    \hat{V}(t) \propto \int_{-\infty}^0& dx\int_{-\infty}^{\infty}dy | \hat{U}_{out}(x,y,z)|^2 \\
    &- \int^{\infty}_0 dx\int_{-\infty}^{\infty}dy | \hat{U}_{out}(x,y,z)|^2.
\end{align*}
Taking into account that there is a relative phase shift among $U_{mn}$ due to Gouy phase, we have (the spectrum of $\hat{V}(t)$ is normalized to the shot noise floor, assuming $\alpha \gg 1$)
\begin{align*}
    \hat{V}(t) &\approx 2F_{00,10}[d\hat{X}_1^{10} \cos\theta_a - ( d\hat{X}_2^{10} - \alpha w_0k\hat{\beta})\sin\theta_a]\\
    &+ 2\sum_{mn\neq 10}( d\hat{X}_1^{mn} \cos\theta_a - d\hat{X}_2^{mn} \sin\theta_a)F_{00,mn},
\end{align*}
where $F_{lm,pq} = \iint dxdy U_{lm}(z=0)U_{pq}^{\ast}(z=0)F_{weight}$ and $F_{weight}$ for split detector is simply $\pm 1$ for left and right half plane.
 
In actual experiment, the light goes through a lens system. The propagation can be done with the help of ABCD matrices~\cite{Siegman1986}. However, Gouy angle calculated this way does not yield the correct phase differences between HG modes. Here, we have replaced Gouy angle $\theta$ with accumulated Gouy angle $\theta_a$ to correctly count for the phase differences between HG modes in lens system~\cite{Erden1997}. Under thin lens approximation the beam curvature get modified but beam waist and phase remain the same which makes Gouy phase fail to describe relative phase difference between HG modes. Accumulated Gouy phase can be calculated by summing up the Gouy phase change during free space propagation between the lenses. In the main text, we just use symbol $\theta$ for simplicity.

The tilting angle $\hat{\beta}$ has 2 sources, one generated by thermal random force $\hat{R}(t)$ and the other generated by optical force $\delta \hat{F}(t)$. When the laser hit at the node of the membrane, because of the vacuum amplitude fluctuation $d\hat{X}_1^{10}$, the membrane experience a fluctuating force~\cite{Pinard1999}(still $\alpha \gg 1$) $$\delta \hat{F}(t) \approx 2\hbar k\iint dxdy | \hat{U}_{in}|^2 \phi(\vec{r}) = 4\hbar k|\alpha|C_{00,10} d\hat{X}_1^{10},$$ where $C_{lm, pq} \equiv\int U_{lm}(z=0)U_{pq}^{\ast}(z=0)\phi(\vec{r})dxdy$ and under small laser beam waist approximation is evaluated to $C_{00,10} = \frac{k_mw_0}{2}$. Then from Eq.~(\ref{EOM}), $$\hat{X}(\omega) = \chi_m(\omega)(\delta \hat{F}(\omega) + \hat{R}(\omega)),$$ and the spectral density of the split photodetector can be calculated to be
\begin{align*}
    S_V(f) &= F_{00,10}^2(\sin^2\theta_a+|\cos\theta_a + 2\hbar D\chi_m\sin\theta_a|^2) \\
    &+ \sum_{mn\neq 10}F_{00,mn}^2+  F_{00,10}^2|2\alpha w_0k\sin\theta_a|^2S_{\beta}^{th}(f)
\end{align*}
where $S_{\beta}^{th}(f) = 2|k_m\chi_m|^2m_{eff}\Gamma k_bT$ (assuming $k_bT\gg \hbar\omega_m$), $S_{\beta}^{zp}(f) = |k_m\chi_m|^2m_{eff}\Gamma\hbar\omega_m$ and $D = |kk_mw_0\alpha|^2$.
And the tilt angle spectral density follows as
\begin{align*}
    S_{\beta}(f) &= \frac{1}{4 N w_0^2 k^2 \sin^2\theta_a} + N\hbar^2 k^2 w_0^2 k_m^4 |\chi_m|^2 + S_{\beta}^{th}(f)\\
    & + \hbar k_m^2 \mathrm{Re}\{\chi_m\}\cot\theta_a + \frac{1}{4 N w_0^2k^2 \sin^2\theta_a}\sum_{mn\neq 10}\frac{F_{00,mn}^2}{F_{00,10}^2}.
\end{align*}
Assuming negligible thermal motion $S_{\beta}^{th}(f) \approx 0$ and perfect detection $\sum_{mn\neq 10}\frac{F_{00,mn}^2}{F_{00,10}^2} = 0$, without considering the correlation term $\hbar k_m^2 \mathrm{Re}\{\chi_m\}\cot\theta_a$, it is easy to see that $S_{\beta}(f)_{min} = \hbar k_m^2 |\chi_m|$ ($\theta = \pi/2$). And taking account of the correlation term $S_{\beta}(f)_{min} = \hbar k_m^2 \mathrm{Re}\{\chi_m\}$ ($\theta$ not at $\pi/2$).

The cooperativity can be derived as SNR of zero point motion to shot noise as
$$ C = \frac{2PQ}{m_{eff}}\frac{kw_0^2k_m^2}{c\omega_m^2}$$

\section{\uppercase{classical noise modulation}}
The calculation for classic noise is similar to that of quantum noise. The input field is $\hat{U}_{in} = \alpha(U_{00} + \frac{\Delta x(t)}{w_0}U_{10})$ assuming large classical noise and $\Delta x(t)$ is the displacement of the laser spot. Following the same procedures, we have
\begin{align*}
    S_{\beta}(f) &= \frac{S_{\Delta x}(f)/w_0^2}{Nw_0^2 k^2\sin^2\theta_a}|\cos\theta_a + 2\hbar D\chi_m\sin\theta_a|^2\\
    &+ S_{\beta}^{th}(f),
\end{align*}

and $S_{\beta}^{Cla}(\theta_a) = \frac{S_{\Delta x}(f)/w_0^2}{Nw_0^2 k^2}\cot^2\theta_a$.


\end{document}


\renewcommand{\theequation}{S\arabic{equation}}
\renewcommand{\thefigure}{S\arabic{figure}}
\renewcommand{\thesection}{S\arabic{section}}

\title{Supplement}
\maketitle{}

\section{Theory}
\subsection{Fourier transformation and power spectral convention}
We use unitary Fourier transformation defined as:
\[h(\omega)=\frac{1}{\sqrt{2\pi}}\int^{\infty}_{-\infty}e^{-i\omega t}h(t)dt\]
\[h(t)=\frac{1}{\sqrt{2\pi}}\int^{\infty}_{-\infty}e^{i\omega t}h(\omega)d\omega\]
A Gaussian white noise $\sigma(t)$ of variant $ \sigma(t)^2  = \bar{\sigma}^2$, its auto-correlation function is $G_{\sigma}(\tau) =  \langle\sigma(0)\sigma(\tau)\rangle = \bar{\sigma}^2\delta(\tau)$.

The spectral density is defined as $$S_h(\omega) = \frac{1}{2\pi}\int_{-\infty}^{\infty}dte^{-i\omega t}G_h(t).$$

Here the energy is conserved in the sense that ($\omega = 2\pi f$) $$\int S_h(f)df = \int S_h(\omega)d\omega = \lim_{t'\to\infty}\frac{1}{t^{\prime}}\int_0^{t^{\prime}} |h(t)|^2dt.$$

\subsection{thermal noise}
According to fluctuation dissipation theorem, the random thermal force $R(t)$ is associated with the dissipation rate $\Gamma$ as $G_R(\tau) = 2m_{eff}\Gamma k_bT\delta(\tau)$. According to Wiener-Khinchin Theorem, the spectral density can be written as $$S_R(\omega) = \frac{1}{2\pi}\int_{-\infty}^{\infty}dte^{-i\omega t}G_R(t) = \frac{2 m_{eff}\Gamma k_bT}{2\pi}$$ and $$S_R(f) = 2\pi S_R(\omega) = 2 m_{eff}\Gamma k_bT$$
It can be shown that it agrees with Equipartition theorem for harmonic oscillator that $\frac{1}{2}k x^2 = \frac{1}{2}k_bT$ $$\braket{x^2} = \int S_x(f)df = \int |\chi_m(f)|^2S_R(f)df = \frac{k_bT}{m_{eff}\omega_m^2}.$$

\subsection{Equation of motion for Membrane}

The membrane is in highly thermal occupied state, so we treat it classically. For the membrane, the motion can be written as $\vec{\hat{\Phi}}(\vec{r}, t) = \hat{X}(t)\vec{\phi}(\vec{r})$ where $\hat{X}(t)$ satisfy equation of motion

\begin{equation}
m_{eff}\ddot{\hat{X}}+m_{eff}\Gamma\dot{\hat{X}}+m_{eff}\omega_m^2\hat{X} = \hat{F},\label{EOM}
\end{equation}

where $m_{eff}$ is the effective mass defined as $m_{eff} = \int d\vec{r}|\vec{\phi}|^2\rho$ ($|\phi_{max}|$ normalized to 1) and $\rho$ is the density. For a membrane of length $a$, $b$ and thickness $h$ and out-of-plane mode shape of $\vec{\phi}(\vec{r}) = sin(\frac{l\pi}{a}(x-\frac{a}{2}))sin(\frac{p\pi}{b}(y-\frac{b}{2}))\frac{\vec{z}}{|z|}$, $m_{eff} = \frac{1}{4}\rho abh$, and $l$ and $p$ are the number of anti-nodes along transverse directions. The response function is $\chi_m(\omega) = \frac{1}{m_{eff}}\frac{1}{(\omega_m^2-\omega^2)+i\omega\Gamma}$ found by solving eqn(\ref{EOM}) in frequency space.

\subsection{Hermite-Gaussian mode convention}
The explicit form of Hermite-Gaussian beam used in the main text is $$U_{l,m}(x,y,z) = \sqrt{\frac{2}{\pi}}\frac{1}{\sqrt{2^{l+m}m!l!w_x w_y}}e^{-\frac{x^2}{w_x^2} - \frac{y^2}{w_y^2}}H_l(\frac{\sqrt{2}x}{w_x})H_m(\frac{\sqrt{2}y}{w_y})e^{-ikz}e^{-ik(\frac{x^2}{2R_{x}(z)} + \frac{y^2}{2R_{y}(z)})}e^{i(l+\frac{1}{2})\theta_x + i(m +\frac{1}{2})\theta_y}$$
$$H_l(x) = (-1)^ne^{x^2}\frac{d^n}{dx^n}e^{-x^2}$$
$$w_{x(y)} = w_{0x(y)}\sqrt{1+(\frac{z}{z_{Rx(y)}})^2}$$
$$R_{x(y)}(z) = z(1+(\frac{z_{Rx(y)}}{z})^2)$$
$$z_{Rx(y)} = \frac{\pi w_{0x(y)}^2}{\lambda}$$
$$tan\theta_{x(y)} = \frac{z}{z_{Rx(y)}}$$
$U_{l,m}(x,y,z)$ defined above has orthonormal reationship as $\int dxdy U_{l,m}(z)U_{p,q}^{\ast}(z) = \delta_{lp}\delta_{mq}$. In most situations we are dealing with $w_{0x} = w_{0y}$, the above equations can be simplified by removing $x,y$ subscripts. 

\subsection{Optical field of the laser}
The input field of a typical laser of power $P$ is $$\alpha U_{00} +\sum_{mn}\hat{\delta}_{mn}U_{mn}. $$ Here $|\alpha|^2=N = \frac{P}{\hbar\omega_L}$. $\hat{\delta}_{mn}$ represents the vacuum fluctuation in all other HG modes, $\hat{\delta}_{mn} = d\hat{X}_1^{mn}+id\hat{X}_2^{mn}$ where $d\hat{X}_1^{mn}$ is the amplitude vacuum fluctuation and $d\hat{X}_2^{mn}$ is the phase vacuum fluctuation. $\langle d\hat{X}_i^{mn}(\omega) (d\hat{X}_j^{pq}(\omega^{\prime}))^{\dagger} \rangle = \frac{1}{4}\delta_{ij}\delta_{mp}\delta_{nq}\delta(\omega - \omega^{\prime})$.

\subsection{Accumulated Gouy phase in lens system}
It is worth noting that the Gouy phase $\theta$ defined above is only valid for free space propagation. For a lens system where the curvature gets modified, we need to replace it with accumulated Gouy phase, which is the sum of Gouy phase change during free space propagations between the lenses~\cite{Erden1997}. Under thin lens approximation, accumulated Gouy phase remain the same going through the lens. 

Another quantity convenient for propagating Hermite-Gaussian beam is complex radius defined as $\frac{1}{q} \equiv  \frac{1}{R(z)} - i\frac{\lambda}{\pi w(z)^2}$. Hermite-Gaussian beam can be uniquely written knowing $q$ whose change through lenses' system can be calculated using ABCD matrices as~\cite{Siegman1986} $$\frac{1}{q} = \frac{A+B/q^{\prime}}{C+D/q^{\prime}}$$. 
$\begin{bmatrix}
A & B\\
C & D
\end{bmatrix}$	 matrix for free space propogation of distance $L$ and thin lens of focal length $f$ are
$\begin{bmatrix}
1 & L\\
0 & 1
\end{bmatrix}$	
and
$\begin{bmatrix}
1 & 0\\
-\frac{1}{f} & 1
\end{bmatrix}$, respectively.

\subsection{Fourier optics calculation}

The reflection of the laser field off the device can be calculated with Fourier optics. The reflected field $\hat{U}_{out}$ can be associated with the input field $\hat{U}_{in}$ as (up to a change of direction of propagation)$$\hat{U}_{out}(z=0) = \hat{U}_{in}e^{-i2k\hat{\Phi}(\vec{r},t)},$$ where $k = \frac{\omega_L}{c}$,  $\hat{\Phi}(\vec{r},t)$ can be approximated to $\hat{\Phi}(\vec{r},t) \approx \hat{X}(t)k_mx$ as the laser beam waist is much smaller than the mechanical wave length ($w_0\ll \frac{2\pi}{k_m}$, $k_{m}$ is mechanical wave numbers) and centered to the anti-node in y direction and the node in the x direction (center of the membrane), where we have dropped $x,y$ footnote and take $k_m$ as $k_{mx}$. 

Based on the recursion relationship of Hermite polynomials of $H_{n+1}(x)=2xH_n(x)-2nH_{n-1}(x)$ one can derive that $$\frac{2}{w}xU_{mn} = \sqrt{m+1}U_{m+1n}+\sqrt{m}U_{m-1n}.$$ With that 

\begin{align*}
    \hat{U}_{out}(z=0) &\approx \hat{U}_{in}(1-i2k\hat{X}(t)k_mx)\\
            &= (\alpha U_{00}+\sum_{mn}\hat{\delta}_{mn}U_{mn})(1-i2k\hat{\beta}(t)x)\\
            &\approx \alpha U_{00}+(d\hat{X}_1^{10} + id\hat{X}_2^{10})U_{10}-\alpha(i2k\hat{\beta}(t)x) U_{00} + \sum_{mn\neq 10}\hat{\delta}_{mn}U_{mn}\\
            &= \alpha U_{00}+(d\hat{X}_1^{10} + id\hat{X}_2^{10})U_{10}-(i\alpha w_0 k\hat{\beta}(t)) U_{10} + \sum_{mn\neq 10}\hat{\delta}_{mn}U_{mn}\\
    \hat{U}_{out}(z=0) &\approx \alpha U_{00}+(d\hat{X}_1^{10}+i(d\hat{X}_2^{10} - \alpha w_0 k \hat{\beta}(t)))U_{10} + \sum_{mn\neq 10}\hat{\delta}_{mn}U_{mn},
\end{align*}

where $\hat{\beta}(t) \equiv \hat{X}(t)k_m$ is the tilting angle at the node.

Propogating the filed out, we can write the field at distance as: $$\hat{U}_{out}(z) = \alpha U_{00}(x,y,z)+(d\hat{X}_1^{10}+i(d\hat{X}_2^{10} - \alpha w_0 k \hat{\beta}(t)))U_{10}(x,y,z)+ \sum_{mn\neq 10}\hat{\delta}_{mn}U_{mn}(x,y,z),.$$

The signal $V(t)$ for the split photodetector at $z$ is proportional to 
\begin{align*}
    \hat{V}(t) \propto \int_{-\infty}^0& dx\int_{-\infty}^{\infty}dy | \hat{U}_{out}(x,y,z)|^2 - \int^{\infty}_0 dx\int_{-\infty}^{\infty}dy | \hat{U}_{out}(x,y,z)|^2.
\end{align*}
Taking into account that there is a relative phase shift between $U_{00}$ and $U_{10}$, we have ($\hat{V}(t)$ is normalized so that the shot noise $ S_V^{shot}  = 1/Hz$) $$\hat{V}(t) \approx 2F_{00,10}[ d\hat{X}_1^{10} cos\theta - ( d\hat{X}_2^{10} - \alpha w_0k\hat{\beta})sin\theta] + 2\sum_{mn\neq 10}( d\hat{X}_1^{mn} cos\theta -  d\hat{X}_2^{mn} sin\theta)F_{00,mn},$$ where $F_{lm,pq} = \int dxdy U_{lm}(z=0)U_{pq}^{\ast}(z=0)F_{weight}$ and $F_{weight}$ for split detector is simply $\pm 1$ for left and right half plane. Also, we have replaced Gouy phase with accumulated Gouy phase but use the same symbol $\theta$.

The tilting angle $\hat{\beta}$ has 2 sources, one generated by thermal random force $\hat{R}(t)$ and the other generated by optical force $\delta \hat{F}(t)$. When the laser hit at the node of the membrane, because of the vacuume amplitude fluctuation $d\hat{X}_1^{10}$, the membrane experience a fluctuating force~\cite{Pinard1999}(still $\alpha \gg 1$) $$\delta \hat{F}(t) \approx 2\hbar k\iint dxdy | \hat{U}_{in}|^2 \phi(\vec{r}) = 4\hbar k|\alpha|C_{00,10} d\hat{X}_1^{10},$$ where $C_{lm, pq} \equiv\int U_{lm}(z=0)U_{pq}^{\ast}(z=0)\phi(\vec{r})dxdy$ and under small laser beam waist approximation is evaluated to $C_{00,10} = \frac{k_mw_0}{2}$. Then from equation (\ref{EOM}), $$\hat{X}(\omega) = \chi_m(\omega)(\delta \hat{F}(\omega) + \hat{R}(\omega)),$$ and the spectral density of the split photodetector can be calculated to be

\begin{equation}
    S_V(\omega) = \frac{F_{00,10}^2}{2\pi}(sin^2\theta+|cos\theta + 2\hbar D\chi_m(\omega)sin\theta|^2) + \frac{1}{2\pi}\sum_{mn\neq 10}F_{00,mn}^2+  F_{00,10}^2|2\alpha w_0ksin\theta|^2S_{\beta}^{th}(\omega) \label{SV}
\end{equation}

where $S_{\beta}^{th}(\omega) = |k_m\chi_m(\omega)|^2S_R(\omega)$ and $D = |kk_mw_0\alpha|^2$. And

$$S_V(f) = F_{00,10}^2(sin^2\theta+|cos\theta + 2\hbar D\chi_m(\omega)sin\theta|^2) + \sum_{mn\neq 10}F_{00,mn}^2+  F_{00,10}^2|2\alpha w_0ksin\theta|^2S_{\beta}^{th}(f)$$

The cooperativity can be derived as SNR of zero point motion to shot noise as
$$ C = \frac{2PQ}{m_{eff}}\times\frac{kw_0^2k_m^2}{c\omega_m^2}$$

\subsection{classic noise modulation}
The calculation for classic noise is similar to that of quantum noise. The input field is $U_{in} = \alpha(U_{00} + \frac{\Delta x(t)}{w_0}U_{10})$ assuming classical displacement noise $\Delta x(t)$ is much larger than the quantum noise, but smaller compared to $w_0$. Following the same procedures, we have
\begin{align*}
    S_V(f) = F_{00,10}^2\frac{4S_{\Delta x}(f)}{w_0^2}|cos\theta + 2\hbar D\chi_m(\omega)sin\theta|^2+ F_{00,10}^2|2\alpha w_0ksin\theta|^2S_{\beta}^{th}(f)
\end{align*}

\section{Experiment}
\subsection{finite beam spot correction}
The mode of interest in the main article has a mechanical wavelength of about 580~$\mathrm{\mu m}$, which is not much larger than the beam waist $w_0$ of the laser at the membrane, which is about 100~$\mathrm{\mu m}$. Two corrections have to be made.

The first correction is the magnitude of the force experienced by the membrane. The calculation above assumes small beam waist can give $C_{00,10}^{\prime} = \frac{1}{2}k_mw_0$. By doing the integration without making a linear approximation, we find the more exact value to be about $C_{00,10} = 86\%\times C_{00,10}^{\prime}$.

The other correction comes from the linear approximation made when reflecting the incoming laser, where $e^{-i2kX(t)sin(k_mx)}\approx(1-i2kX(t)k_mx)$. When the beam waist is comparable to the mechanical wave length, a better approximation is $e^{-i2kX(t)sin(k_mx)}\approx(1-i2kX(t)sin(k_mx))$. Instead of decomposing the reflected field into different orders of Hermite-Gaussian modes, which require using recursion relationships, we approximate the field on the photodetector by using the fact that the field $g(x, y)$ at the Fourier plane is just a 2D Fourier transformation of the object field $U_{out}(x',y')$. Since all of our experiment data is taking near Fourier plane, we can apply this correction independent of quadrature angle for $\theta \sim 90^{\circ}$. Explicitly, $$g(x,y) = \frac{i}{\lambda f}e^{-i2kf}\int dx^{\prime}dy^{\prime}U_{out}(x',y')e^{-ik(\frac{x}{f}x' + \frac{y}{f}y')}$$
$$g(x,y) \propto e^{-(\frac{x^2+ y^2}{w^2})}\cdot (1 + 2 k Xsinh(k_mw_0\frac{x}{w})e^{-(\frac{k_mw_0}{2})^2}).$$ Here, $w$ refers to the beam spot size at the Fourier Plane and $f$ is the focal length of the output lens. With finite beam spot size $w_0$, the exponential term $e^{-(\frac{k_mw_0}{2})^2}$ leads to a decrease of the effective $D$ defined above to 55\%. 

Such correction is negligible for the torsional mode of the string device characterized in Fig.~4(b) as the tilting at the center is a good approximation to a rigid lever.  The laser beam waist $w_0$ in this case is about $3.2~\mathrm{\mu m}$ and the string width at the center is about $10~\mathrm{\mu m}$ (about $2~\mathrm{mm}$ long and $120 \mathrm{nm}$ thick). We also note that as shown in Fig.~4(b) our string has periodically modulated width that creates a phononic crystal shield for higher order out-of-plane and torsional mode. This phononic shield does not affect the mode shape of the fundamental mode and is irrelevant for the data we presented.
\subsection{Experiment Calibration}

In the experiment, the laser spot position is modulated with an AOM. However, the crystal introduces aberrations to the Gaussian beam. The major aberration is the beam waist size. At the split photodetector it is calculated to be about 520~$\mathrm{\mu m}$ but measured to be about 590~$\mathrm{\mu m}$. This inevitably lowers the absolute magnitude of the signal $V(t)$ by a factor of $\frac{520}{590} \approx 0.88$. This factor is taken into account by artificially multiplying the calculated value of $V(t)$ by this number. The following calibrations are also based on the measured 590~$\mathrm{\mu m}$.

The most important quantity that needs to be calibrated in back action evasion experiment is the actual beam spot displacement amplitude at the device, $\Delta x_0$, when we drive the AOM, as it determines the back action motion amplitude. To get this number right, we took 2 approaches which yield similar results. 

In the first approach, we calibrate $\Delta x_0$ by measuring the resulting beam angle near the Fourier plane by scanning our split photodetector with a calibrated translation stage along the beam path. According to the manufacturer's spec sheet of the AOM, the Bragg angle should be 10~mrad with 80~MHz carrier frequency, which corresponds to $\Delta x_0$ of about 31~$\mathrm{\mu m/MHz}$. We drive the AOM at the specified carrier frequency with an FM deviation of 1~MHz whose modulation frequency is low at 100~Hz so that we can directly observe the sinusoidal wave amplitude at the split photodetector. The above yields a beam angle at the Fourier plane of 0.2~mrad. Knowing the output lens focal length, we infer $\Delta x_0$ to be about 36~$\mathrm{\mu m/MHz}$. Then we correct for the AOM responses to different FM frequencies and amplitudes, by measuring the relative response from the split photodetector. We estimate $\Delta x_0$ to be 2.9~$\mathrm{\mu m}$ for the experimental conditions for the data presented in the main text Fig.~3.

In the second approach, we calculate $\Delta x_0$ from the power spectrum measured by the split photodetector during chirp modulation. By comparing to the thermal motion power spectrum, we obtained the absolute amplitude of optically driven motion. From our calculated effective mass, we know how much force is optically applied and $\Delta x_0$ is found also to be about 2.9~$\mathrm{\mu m}$. This is the number used in the main text.

\subsection{thermal stabilization}

\begin{figure}[h]
\includegraphics[scale = 0.18]{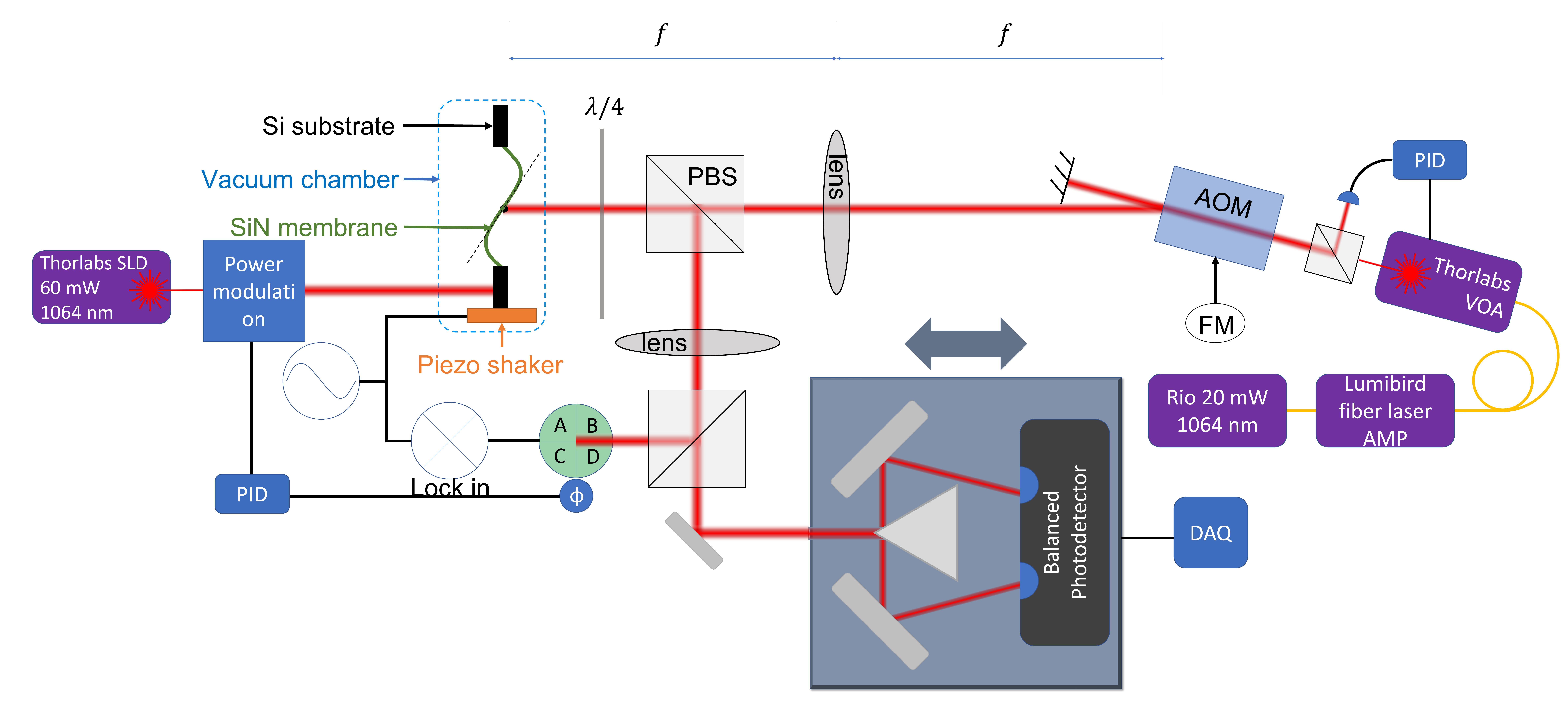}
\caption{\label{fig:DExp} Detailed experiment scheme}
\end{figure}

As the temperature alters the stress of the SiN membrane due to thermal expansion, drift of temperature shifts the resonance frequency of all mechanical modes by a same ratio. To stabilize temperature of the device, we implement 3 measures. 

Firstly, we put a TEC element underneath the sample holder and attach a thermometer on top of the sample holder. Copper ribbons are used to dissipate heat from the back of the TEC element to the vacuum chamber directly (the translation stage in between make the thermal conductance extremely small). A  PID controller is used to stabilize the temperature of the sample holder. 

Secondly, the probe laser power is stabilized with an electronic variable optical attenuator (VOA) where the small amount of light is collected to monitor the laser power fluctuation and used as input for an FPGA based PID controller. 

Thirdly, we directly stabilize the frequency drift by heating the device substrate with an LED. A function generator is used to drive a chosen mechanical mode (different from the mode inspected in the main text) via the piezo shaker attached underneath the sample holder with a sinusoidal wave at a fixed frequency near the resonance. The response obtained by feeding a small amount of output beam (less than 1\%)into a quadrant photodetector is mixed with the drive signal to extract phase information. The phase information is then used as the input for a PID loop which adjusts the amount of LED power hitting at the back of the device substrate, which thermally actuates the resonance frequency. The LED power modulation is done by doing amplitude modulation to an AOM. With the methods above, we are able to stabilize the resonance frequency of the mode we are measuring in the main text at sub-Hz level for days. 

\subsection{mode identification}
The mode frequency can be estimated as

\begin{equation}
    \omega = c\sqrt{(\frac{l\pi}{a})^2+(\frac{p\pi}{b})^2}\label{Rec}
\end{equation}

\noindent where $c$ is the speed of the sound, $a$ and $b$ are the membrane side lengths, and $l$ and $p$ are the number of anti-nodes along transverse directions. We extract a number of resonance peaks observed in the spectrum, fit their locations with the model by adjusting the exact sound of speed only, and find an excellent agreement. Thus, we are confident what spacial mode a specific peak correspond to. To further confirm our mode identification, we set up a dark field imaging system and drive the device with the piezo shaker strong enough to observe the mode shape profile directly. We found some of the modes hybridized, but the mode inspected ($l = 12, p = 3$) in the main text is as expected. 

\subsection{ring down}
To measure the quality factor of the mode inspected in the main text, we performed a ring down measurement using the piezoelectric actuator clamped underneath the sample holder. 

\begin{figure}[h]
\includegraphics[scale = 0.5]{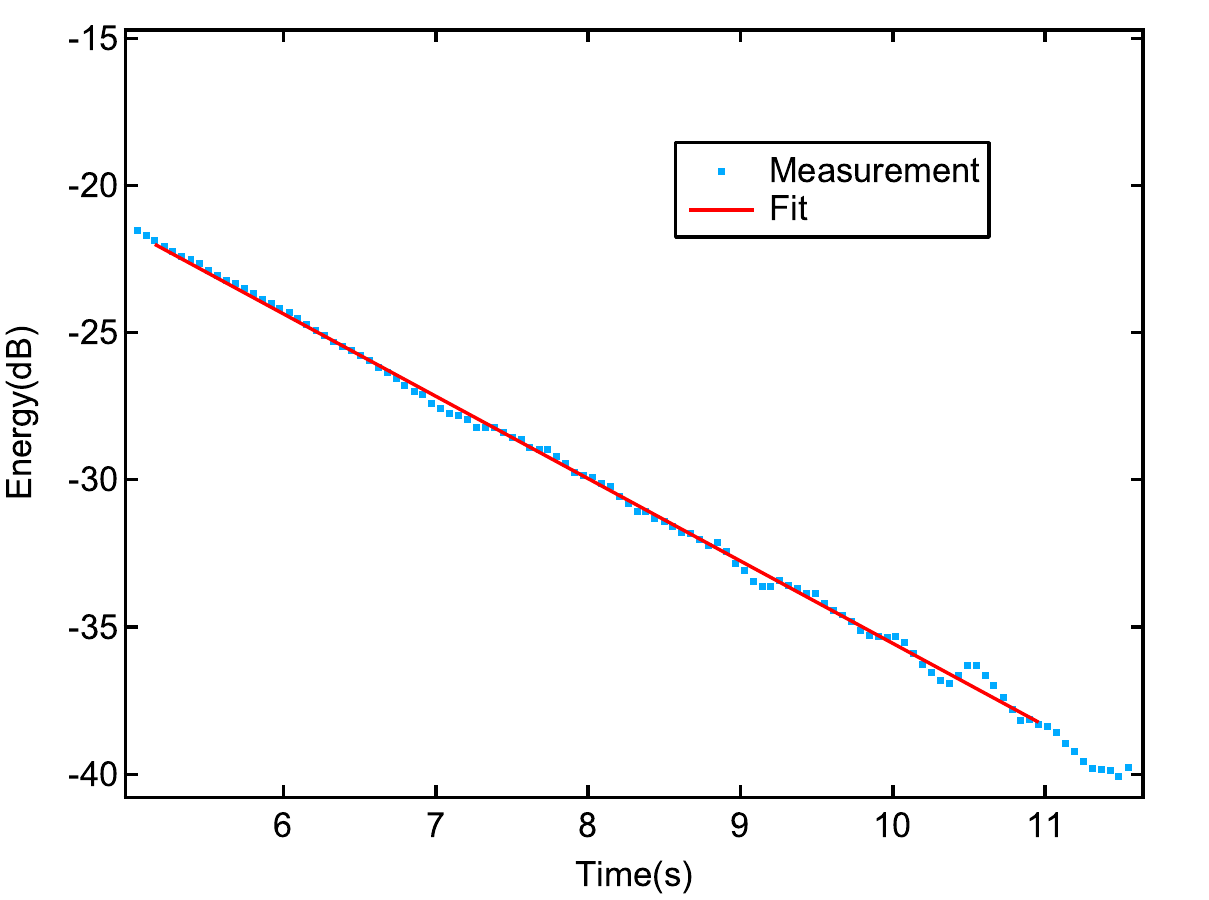}
\caption{\label{fig:HGexpansion} Ring down of the mode inspected in the main text($fm = 371696 Hz$). $Q \approx 3.2\times 10^6$}
\end{figure}

\subsection{Details regarding quadrature angles}
As shown in Fig.~\ref{fig:DExp}, a movable split photodetector is in practice implemented by translating a breadboard that holds a knife-edge prism and a balanced photodetector (Newport 2117-FC) which has high common mode rejection ratio and wide bandwidth. We also include an additional set of relay lenses between the output lens and the knife-edge mirror for convenience, which slightly magnifies the beam and adds about an extra $-180^{\circ}$ phase, placing the Fourier plane at $-270^{\circ}$, near our knife-edge prism. We calculate the accumulated Gouy angle based on ABCD matrices using the measured distances between optical elements. The calculated location of the Fourier plane is within $\sim 2$~mm (corresponds to about $4^{\circ}$ of Gouy angle) of the empirically measured location found by comparing to our model of the optical lever response. The actual phase angle at different positions reported in the main text is based on the ABCD matrices calculation, but globally shifted (by $4^{\circ}$) so that the curve has no back action evasion corresponds to $-270^{\circ}$ (or $90^{\circ}$ reported in the main text). We ascribe such error to the uncertainty in measuring the physical locations of the optical elements, and in a small discrepancy of the exact location of the laser focus relative to the device plane, which translates to large Gouy angle difference. 

To get the data shown in main text, we translate the breadboard back and forward 14 times, spanning roughly 2 days, sampling more than 200 positions. The translation range is about 11 mm. Theoretically, with fixed experimental parameters, the quadrature angle is only a function of position of the breadboard. However, in reality the quadrature angle can drift slowly in time, which we think is due to environment changes such as thermal expansion. In this case, instead of directly converting the translation stage position to angle, we recalibrate them by referencing the quadrature angle to the chirp floor background level, which is minimized at $\theta = 90^{\circ}$. The correction is small, but allows us to combine spectra from each run. 

The chirp signal, created by an arbitrary waveform generator (AWG), is 5~s long and constantly repeats for the entire data run (two days). The trigger signal from the AWG is sent to the data acquisition (DAQ) system (PicoScope 5444D) for syncing. The data acquisition start with sync signal and sampled for the same amount of time of the chirp. Then 10 independently (not continuously, roughly spanning 2 to 3 minutes) taken data are averaged to one trace as shown in main text Fig.~3(c).

The spectrum of the chirp function is not perfectly flat. To get a cleaner response spectrum of the optical lever system, we normalize our spectrum to an identical chirp applied to the AOM only a few kHz shifted above the resonance frequency, where there is no mechanical response and the spectrum is originally just shot noise. This also makes the result robust against noise sources such as small amplitude fluctuation of the laser.

\subsection{Laser heating of the string device}
The string device characterized in the main text as shown in Fig.~4(b) is cooled by a cryostat to the base temperature of 5~K. When we focus the laser to the string, we expect the string to heat up due to optical absorption. To estimate the mode temperature, we model the spectrum using a combination of measured and calculated system parameters. We estimate the beam waist at the device to be $4\pm 1~\mathrm{\mu m}$ based on the manufacturer's specified parameters of lenses and input optical fiber. The effective mass of the fundamental mode is estimated based on the mask dimension, SiN growth thickness and density of $\rho \approx 3100 ~\mathrm{kg/m^3}$ and is $7.3 \pm 0.36 \times 10^{-13} ~\mathrm{kg}$. The laser power of $12.2~\mathrm{mW}$ is measured in the front of the knife-edge mirror, and we estimate the power actually collected by the 2 ports of the balanced photodetector to be more than 85\%. Based on these estimations, we calculate the theoretical spectrum and compared with the experiment result, and find the mode temperature to be about $20\pm10~\mathrm{K}$.

\begin{figure}[h]
\includegraphics[scale = 0.7]{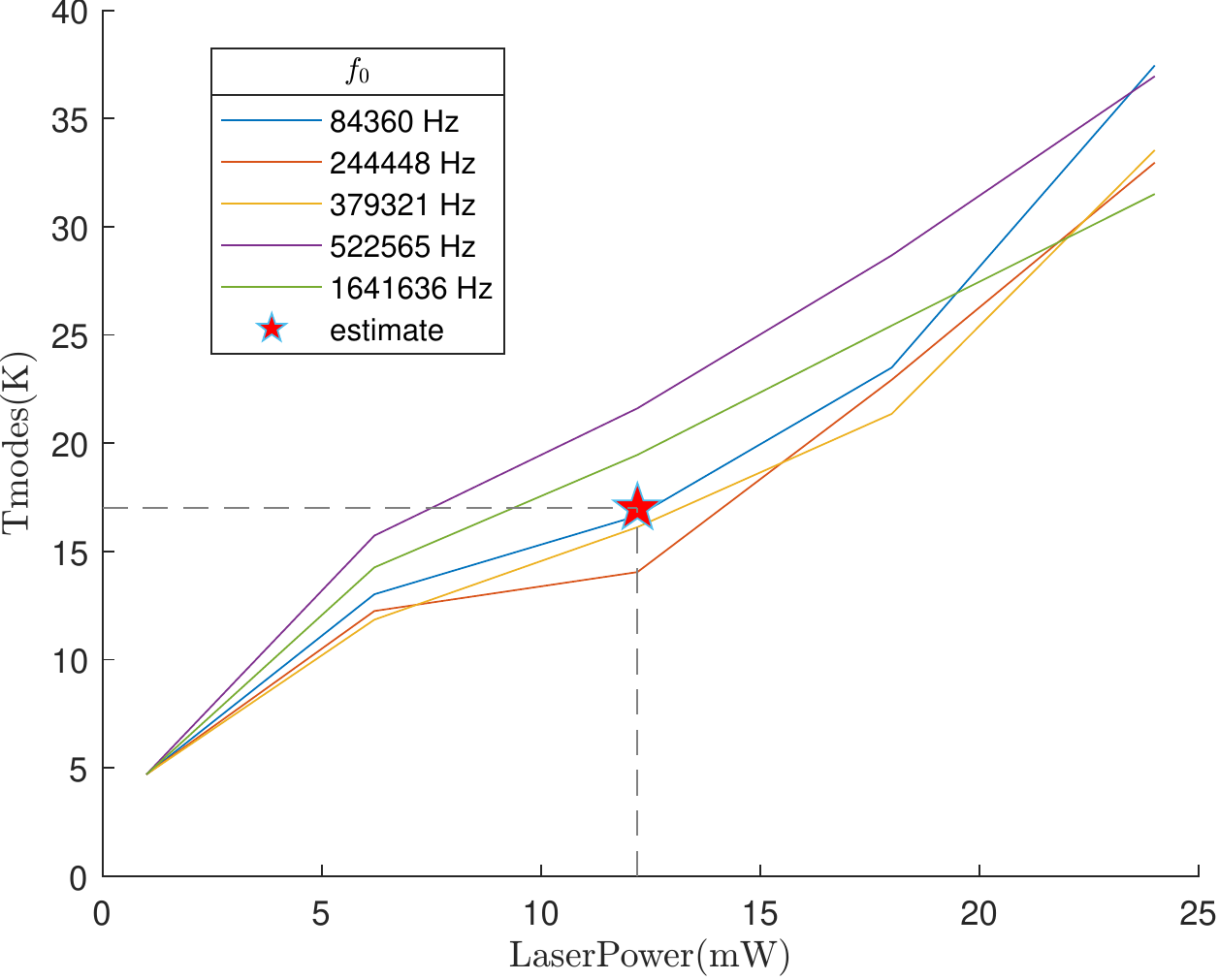}
\caption{\label{fig:LaserHeatingInt} Mode temperature versus laser power. The mode temperature is obtained by normalizing the integral of the thermal motion spectrum to that at the probing laser power of 1~mW, where the device is at the base temperature of the cryostat of 4.8~K. As the thermal motion spectrum integral is proportional to temperature, we can estimate the mode temperature at under different laser powers. Different colors represent different torsional modes.}
\end{figure}

To confirm our estimation of the mode temperature, we also estimate the temperature of several torsional modes by comparing the integral of their thermal motion spectrum (which is proportional to temperature) at different laser power. The integral is normalized to the integral at a low laser power of 1~mW, where the device is at around the base temperature of the cryostat of 4.8~K, and termed as "Tmodes" in the Fig.~\ref{fig:LaserHeatingInt}. As can be seen the mode temperature is roughly linear with regard to laser power, and at the laser power of 12.2~mW the mode temperature is around 17~K, supporting our estimation of temperature.

\begin{figure}[h]
\includegraphics[scale = 0.6]{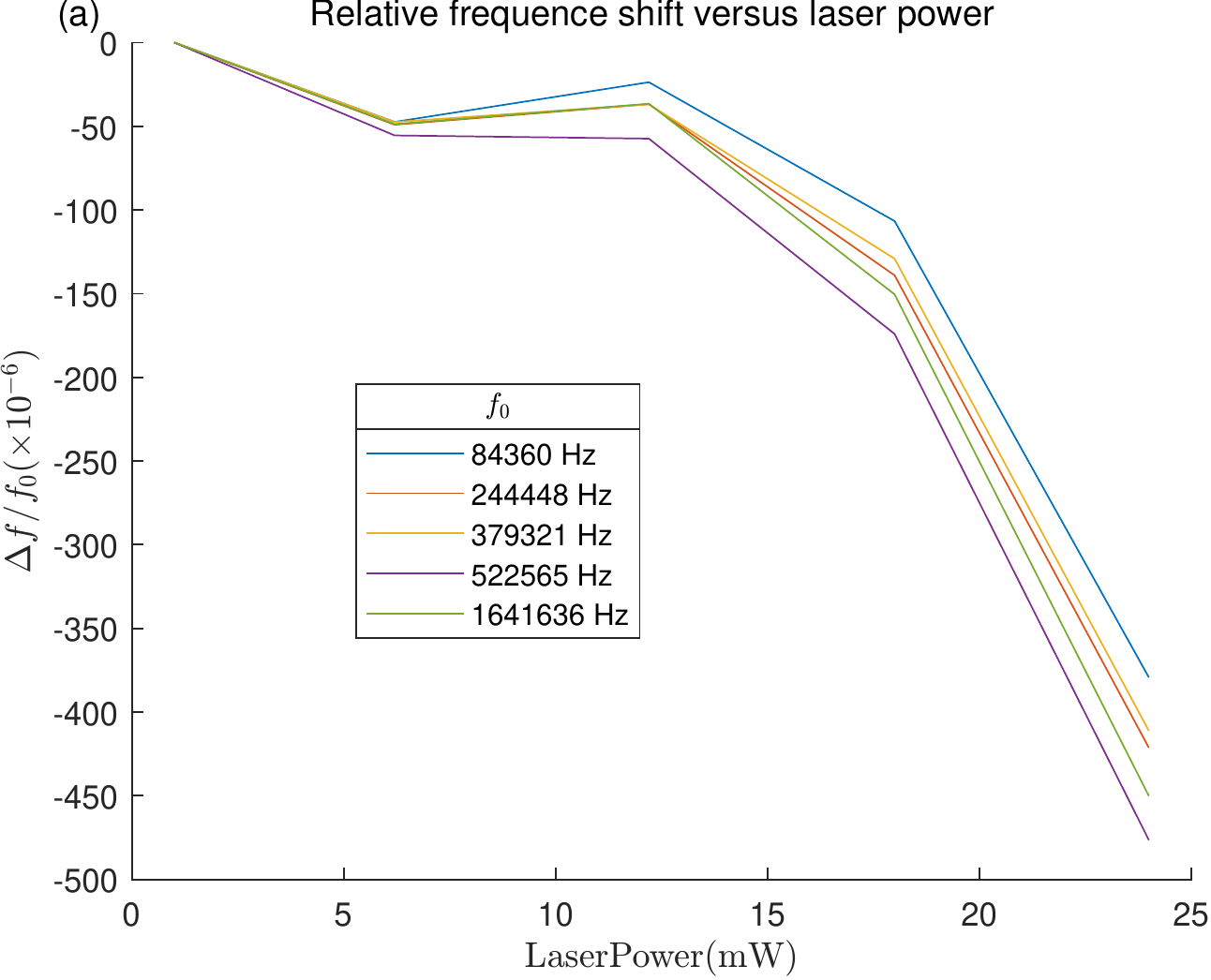}
\includegraphics[scale = 0.6]{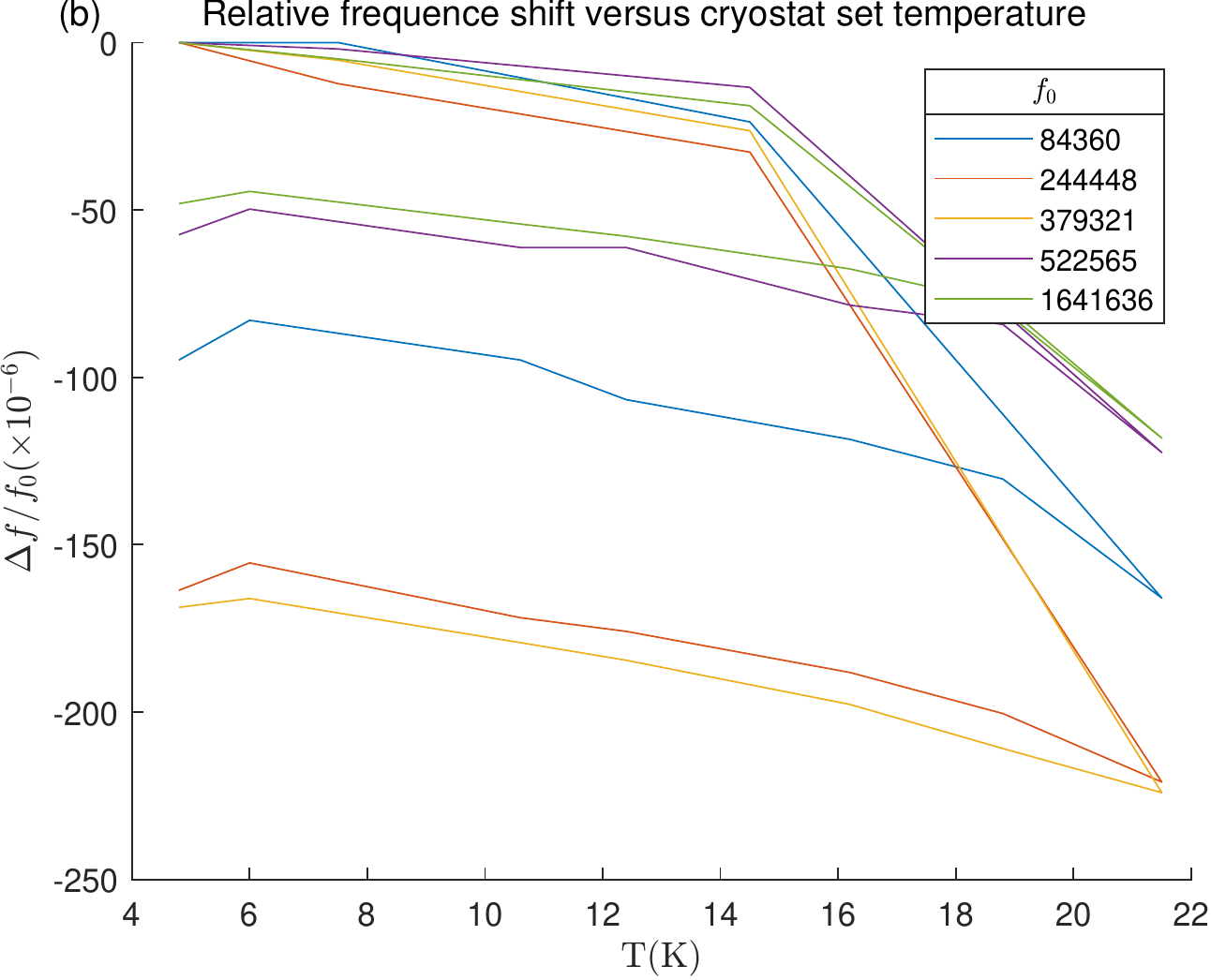}
\caption{\label{fig:LaserHeating} Relative frequency shift versus laser power and cryostat set temperature. $\mathrm{f_0}$ are the resonance frequencies of the modes under probe laser of 1~mW and cryostat set temperature of 4.8~K. As shown in (a), the shifts of the resonance frequencies of different modes are common, indicating that the drifts are due to global factors such as stress or temperature. When the laser power increases to more than 15~mW, we observe more drastic change of resonance frequencies. We perform a rough linear fit for the data below 15~mW, which gives a fractional frequency shift of about $65\times 10^{-6}$ at 12.2~mW. In (b), we cycle the cryostat set temperature from 4.8~K to about 22~K and observe again a large shift above 15~K, together with a permanent shift after cooling back down. Despite this hysteric behavior, the slope below 15~K for the heating and cooling process are comparable and are about $3.3\times 10^{-6}/\mathrm{K}$. Together, we estimate the temperature of the device with probe laser of 12.2~mW as used in the main text is around 25~K.}
\end{figure}

As a third method to estimate the temperature of the device, we look at the temperature dependence of the mode frequencies. We measure the relative frequency shifts versus laser power for several torsional modes, as shown in Fig.~\ref{fig:LaserHeating}(a). We also independently measure the relative resonance frequency shifts at different cryostat temperatures as measured by a silicon diode thermometer, using low optical power, as shown in Fig.~\ref{fig:LaserHeating}(b). By comparing the relative resonance frequency shifts over the range where data is roughly linear, we can estimate the temperature of the device.
We find the actual temperature of the device under laser of $P = 12.2~\mathrm{mW}$ to be at about 25~K. To further confirm our estimation, we observe several torsional modes.

All three methods are consistent with the estimate of mode temperature of 20~K. We also note that for our membrane device at room temperature, we observe no significant self-heating effect due to the laser.

\bibliography{supplement}